\documentclass{article}
\usepackage{graphicx}
\graphicspath{%
    {converted_graphics/}
    {C:/1KGLaaa/aatempfold/Papers/Paper5-PrismSig0/}
}
\begin{document}

\begin{center}
{\LARGE Crystal Growth in the Presence of Surface Melting:} \vskip6pt

{\LARGE Novel Behavior of the Principal Facets of Ice}\vskip6pt

{\Large K. G. Libbrecht and M. E. Rickerby}\vskip4pt

{\large Department of Physics, California Institute of Technology}\vskip-1pt

{\large Pasadena, California 91125}\vskip-1pt

\vskip18pt

\hrule\vskip1pt \hrule\vskip14pt
\end{center}

\textbf{Abstract}. We present measurements of the growth rates of the
principal facet surfaces of ice from water vapor as a function of
supersaturation over the temperature range $-2$ $\geq T\geq -40$ C. Our data
are well described by a dislocation-free layer-nucleation model,
parameterized by the attachment coefficient as a function of supersaturation 
$\alpha (\sigma )=A\exp (-\sigma _{0}/\sigma ).$ The measured parameters $%
A(T)$ and $\sigma _{0}(T)$ for the basal and prism facets exhibit a complex
behavior that likely originates from structural changes in the ice surface
with temperature, in particular the onset and development of surface melting
for $T>-15$ C. From $\sigma _{0}(T)$ we extract the terrace step energy $%
\beta (T)$ as a function of temperature for both facet surfaces. As a basic
property of the equilibrium ice surface, the step energy $\beta (T)$ may be
amenable to calculation using molecular dynamics simulations, potentially
yielding new insights into the enigmatic surface structure of ice near the
triple point.

\section{Introduction}

Surface melting occurs when the near-surface atomic or molecular layers of a
crystalline solid are not as tightly bound as the deeper layers, causing the
near-surface layers to lose their ordered structure (for a review see \cite%
{surfacemelting}). The result is an amorphous \textquotedblleft
premelted\textquotedblright\ layer, also called a quasi-liquid layer, that
exists in equilibrium at the solid surface. The structure of the
quasi-liquid layer is strongly temperature dependent, and its thickness
typically diverges as the melting point is approached. Surface melting is a
common phenomenon in metals and other simple crystalline materials, and it
has been especially well studied in ice \cite{dash, wei, dosch}.

In general the effects of surface melting on crystal growth have been little
explored, although clearly the structural changes associated with surface
melting can have a profound effect on surface molecular dynamical processes 
\cite{smxtalgrowth1, smxtalgrowth2}. Since our theoretical understanding of
surface melting is relatively poor, we sought to examine how changes in
surface structure with temperature affect crystal growth behavior.

It has long been long suspected that surface melting plays an important role
in the growth dynamics of ice crystals from water vapor \cite{kurodalac,
kkreview, libbrechtreview}. Although ice is a monomolecular crystal with a
simple hexagonal structure under normal atmospheric conditions, ice crystals
forming from water vapor exhibit an exceedingly rich spectrum of plate-like
and columnar morphologies as a function of temperature and supersaturation
over the temperature range $0$ $\geq T\geq -40$ C \cite{libbrechtreview,
morph1, morph2}. Since the premelted layer in ice develops over this same
temperature range \cite{dash, wei, dosch}, the prevailing thinking holds
that the temperature-dependent effects of surface melting on ice crystal
growth are responsible for the observed morphological complexities, together
with instabilities arising from diffusion-limited growth and other effects 
\cite{libbrechtreview}. To date, however, this long-held hypothesis has
remained largely untested by quantitative experimental data.

As part of our investigation, we made precise measurements of the growth
rates of small faceted ice crystals from water vapor under carefully
controlled conditions, in order to better quantify and parameterize the
intrinsic ice growth behavior. To this end we measured growth rates of the
basal (0001) and prism (\={1}100) ice surfaces as a function of water vapor
supersaturation $\sigma $ and temperature $T$ over the temperature range $-2$
$\geq T\geq -40$ C, thus covering the onset and development of surface
melting. Our measurements were made at low background pressure to reduce the
effects of particle diffusion through the surrounding gas, so the growth was
mainly limited by surface attachment kinetics.

\begin{figure}[t] 
  \centering
  \includegraphics[bb=0 0 1234 1130,width=3.75in,height=3.43in,keepaspectratio]{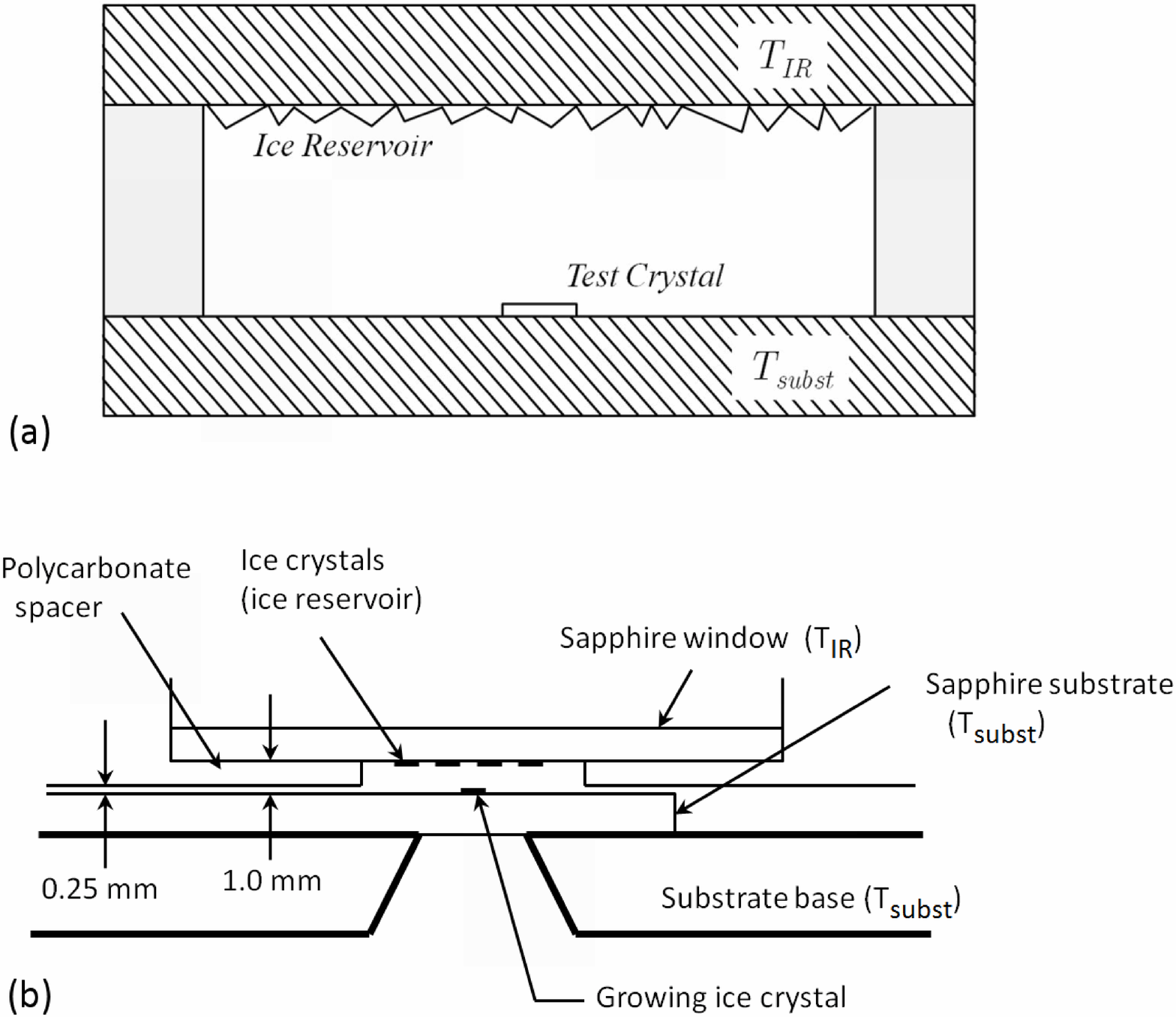}
  \caption{(a) An idealized schematic of
the inner sub-chamber of our experimental set-up. The top surface, covered
with ice crystals, acts as a water vapor reservoir at temperature $T_{IR}$
for a growing test crystal resting on the substrate at temperature $%
T_{subst}.$ When $T_{IR}>T_{subst}$, growth rates are determined by
measuring the thickness of the crystal (i.e. the distance between the
substrate and the parallel top facet) as a function ot time using optical
interferometry, and by measuring the other crystal dimensions using optical
microscopy viewing from below the substrate. (b) A rough schematic of the
actual test chamber, located inside a larger vacuum chamber. The substrate
can be rotated to bring test crystals into position under the ice reservoir.
Additional experimental details are provided in \protect\cite{details}.}
  \label{basic}
\end{figure}

\section{Ice Crystal Growth Measurements}

The goal of our ice growth experiments was to examine the growth of
individual ice crystal facets in a carefully controlled environment, and an
idealized schematic diagram of our experimental set-up is shown in Figure %
\ref{basic}a. The top surface of the experimental chamber consists of a
thermal conductor at a uniform temperature $T_{IR},$ and a layer of ice
crystals on its lower surface serves as a source of water vapor. At the
beginning of each measurement, a single test crystal is placed near the
center of the bottom substrate surface held at temperature $T_{subst}$. The
ice reservoir and the substrate are separated by thermally insulating walls
with a vertical spacing of 1.0 mm. The temperature difference $\Delta
T=T_{IR}-T_{subst}$ determines the effective water vapor supersaturation
seen by the test crystal.

During a typical experimental run, we continuously nucleated ice crystals in
a much larger outer chamber containing ordinary laboratory air (see \cite%
{details}), where they grew while slowly falling to the bottom of the
chamber. Typically $>10^{7}$ of these micron-scale crystals were growing
within the outer chamber at any given time, and the fall times were
approximately 3-5 minutes. This cloud of slowly growing crystals served as
the source of seed crystals for our growth measurements.

To select a test crystal, an inlet valve on the top of a smaller inner
chamber was opened, and air carrying suspended ice crystals was drawn from
the outer chamber through the inner chamber. The operator rotated the
substrate while observing the test region under the ice reservoir (see
Figure \ref{basic}), thus examining crystals that randomly landed on the
substrate. When a suitable test crystal was identified, the inlet valve was
closed and the pressure in the inner chamber was reduced to $<30$ mbar.

Once the pressure was stable, the operator first adjusted $T_{IR}$ and
observed the test crystal growing or evaporating slightly in order to
determine the $\sigma =0$ point, which typically took a few minutes. After
this, $\Delta T$ was slowly increased to grow the test crystal. The
thickness of the crystal -- defined as the distance between the substrate
and the parallel top facet -- was determined using optical interferometry,
while optical imaging was used to record the crystal size and morphology in
the substrate plane. These data, along with the temperature difference $%
\Delta T,$ were all recorded as the crystal grew. After a few minutes, when
the overall crystal size exceeded $\sim 100$ $\mu $m, the crystal was
evaporated away along with any other crystals on the substrate, and another
crystal was selected.

\begin{figure}[h] 
  \centering
  \includegraphics[width=3.7in,keepaspectratio]{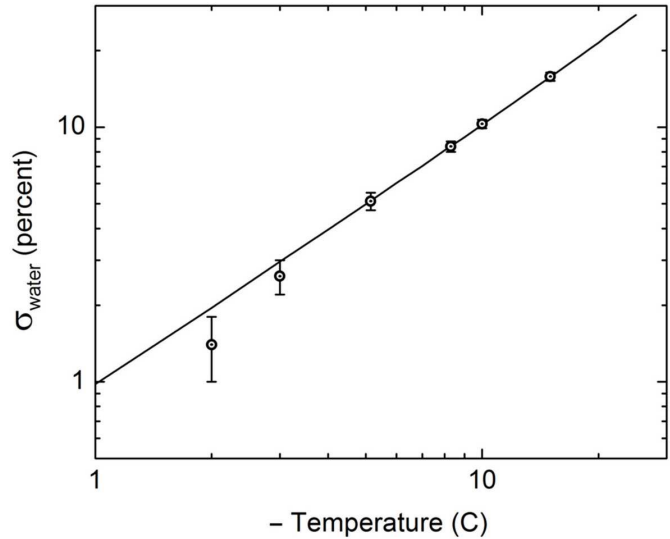}
  \caption{Measurements of the
supersaturation of liquid water with respect to ice in our test chamber,
determined by nucleating and observing water droplets on the substrate. The
line gives the accepted $\protect\sigma _{water}(T)$ from \protect\cite%
{mason}.}
  \label{watersupersat}
\end{figure}

One important aspect of our experiment was the accuracy of the water vapor
supersaturation $\sigma $ seen by each test crystal. Our typical precision
in setting $\Delta T$ was $\pm 0.003$ C, and drifts in $\Delta T$ occurring
during a typical measurement were comparable. This overall temperature
uncertainty corresponded to an uncertainty in the $\sigma =0$ point for each
crystal of $\Delta \sigma \approx \pm 0.03$ percent. If extra care was taken
to determine the $\sigma =0$ point and stabilize the temperature at the
initial stage of a growth run, then this uncertainty could be reduced
perhaps a factor of two further.

We checked our calculated supersaturation, based on the measured $\Delta T,$
by nucleating water droplets on the substrate (in the absence of any test
crystals) and measuring $\Delta T$ at which the droplets were neither
evaporating nor growing. From this we extracted the supersaturation of water
relative to ice $\sigma _{water}$ as a function of temperature, shown in
Figure \ref{watersupersat}. Accurately determining the droplet stability
point became difficult at the higher temperatures, and this difficultly
likely explains the systematic trend away from $\sigma _{water}(T)$ seen in
Figure \ref{watersupersat}. The fact that our measurements, using no free
parameters, were in good agreement with the accepted $\sigma _{water}(T)$
suggests that the water vapor supersaturation was quite well known in these
experiments.

Only crystal facets parallel to, and thus not in contact with, the substrate
were used to determine the intrinsic crystal growth parameters. Facets that
intersected the substrate often grew at somewhat higher rates, especially at
low $\sigma ,$ owing to substrate interactions that reduced the normal
nucleation barriers on these facets \cite{substrateinteractions}.

Broad-band interferometry was used when measuring the basal facets, as
described in \cite{details, previous}, allowing an absolute measurement of
the crystal thickness. This technique worked well for thin plate-like
crystals, but was less effective when measuring the growth of the prism
facets, owing to the greater distance between the top facet and the
substrate. Thus all the prism facet measurements were taken using laser
interferometry as described in \cite{oldgrowth}.

\begin{figure}[ht] 
  \centering
  \includegraphics[width=2.3in,height=4.39in,keepaspectratio]{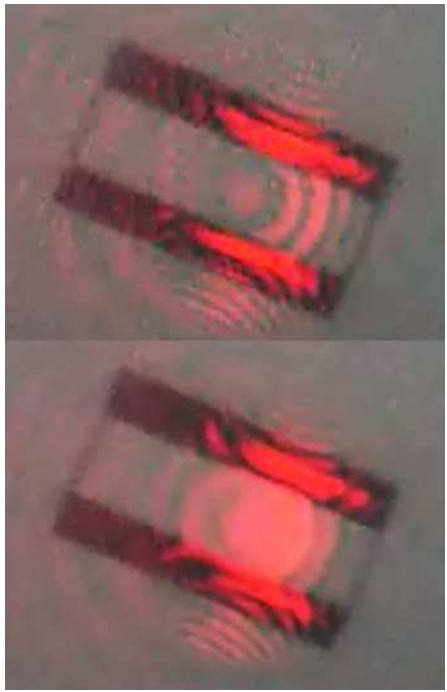}
  \caption{Two images of a single ice prism
taken at different times. Interference between reflections from the
substrate/ice and ice/vacuum interfaces produced the central laser spot seen
in these images. The brightness oscillated between dark (top image) and
bright (lower image) as the crystal grew. The c-axis length of this prism is
62 $\protect\mu $m.}
  \label{prisms}
\end{figure}

For example, Figure \ref{prisms} shows two images of a single ice prism
taken at different times. The central laser spot oscillated between dark and
bright as the crystal thickness increased, resulting from the interference
of reflections from the sapphire/ice and ice/vacuum interfaces. These
brightness changes were used to measure the growth velocity of the top prism
facet. For these data the initial crystal morphology was assumed to be a
simple hexagonal prism, so imaging of the crystal yielded an estimate of the
initial crystal thickness.

While our crystal size measurements were straightforward in most cases \cite%
{details}, we did run into some initial problems using laser interferometry
on smaller crystals. In addition to the central \textquotedblleft
bulls-eye\textquotedblright\ fringes seen in Figure \ref{prisms}, one can
also see a number of smaller laser fringes that we have come to calling
\textquotedblleft false\textquotedblright\ fringes, resulting from multiple
reflections within the crystal. The false fringes tend to dominate the laser
interference pattern when the top facet is small, and the resulting
bright/dark transitions are not always obvious to interpret. We came to
realize that in an earlier version of this experiment \cite{oldgrowth}, we
sometimes inadvertently used the false fringes when measuring prism facet
growth, thus leading to erroneous results. The basal growth measurements in 
\cite{oldgrowth} were not compromised by this problem, since the basal facet
surfaces were large enough to produce clean laser fringes.

Some other important aspects of the experiment included: 1) Our test
crystals were small, typically $<100$ $\mu $m in overall size. Thin plates
were typically $<5$ $\mu $m thick. The background air pressure in the test
chamber was typically $<30$ mbar. Using small crystals and low background
pressures was important to reduce the effects of particle diffusion, as
described further below; 2) We used only crystals with simple morphologies
and well formed facets, and each crystal was discarded after growth.
Evaporating and regrowing crystals was found to result in generally lower
quality data; 3) Low outgassing materials were used to construct our
experimental chamber, and the thermal control hardware (including
thermoelectric modules, thermally conducting grease, and associated wiring)
was all mounted outside the vacuum envelope. The chamber was also baked
between each run to remove any volatile chemical residues. We believe that
chemical impurities were responsible for some discrepancies in our earlier
results \cite{oldgrowth}, and these problems have been remedied for the
current experiment; 4) Our test crystals were freshly made in a clean
environment and transported within minutes to our test chamber with minimal
processing, as described above. This also helped minimize any buildup of
chemical impurities on the ice surfaces. With all these precautions, we
believe that we have adequately reduced many systematic errors that appear
to have been present in previous ice growth experiments \cite%
{libbrechtreview, critical}.

The goal of our measurements was to determine the intrinsic growth rates of
the principal facets of ice, which we write in terms of the surface
attachment coefficient $\alpha _{instrinsic}$ using $v=\alpha
_{intrinsic}v_{kin}\sigma _{surf},$ where $v$ is the growth velocity normal
to the surface, $v_{kin}(T)$ is a kinetic velocity derived from statistical
mechanics \cite{libbrechtreview}, and $\sigma _{surface}$ is the water vapor
supersaturation immediately above the growing ice surface. The maximum
allowed growth velocity has $\alpha _{intrinsic}=1$ (albeit with some
caveats; see \cite{libbrechtreview}), while faceted surfaces generally have $%
\alpha _{intrinsic}<1.$

We also found it useful to define a \textquotedblleft
measured\textquotedblright\ attachment coefficient $\alpha _{meas}$ derived
entirely from experimentally measured quantities using $v=\alpha
_{meas}v_{kin}\sigma _{\infty },$ where $\sigma _{\infty }$ is the
supersaturation far from the crystal. To lowest order, the measured $\alpha
_{meas}\left( \sigma _{\infty }\right) $ is given by 
\begin{equation}
\alpha _{meas}\left( \sigma _{\infty }\right) \approx \frac{\alpha
_{intrinsic}\left( \sigma _{\infty }\right) \alpha _{diff}}{\alpha
_{intrinsic}\left( \sigma _{\infty }\right) +\alpha _{diff}}  \label{fitform}
\end{equation}%
as described in \cite{libbrechtreview, details}, where $\alpha
_{diff}\approx 0.15(R_{1}/R_{eff})(P_{1}/P)$, $R_{eff}$ is an effective
crystal radius, $P$ is the background air pressure, $R_{1}=1$ $\mu $m, and $%
P_{1}=1$ bar. If the measured crystal growth is predominantly kinetics
limited, then $\sigma _{surf}\approx \sigma _{\infty }$ and $\alpha
_{meas}\left( \sigma _{\infty }\right) \approx \alpha _{intrinsic}\left(
\sigma _{surf}\right) .$ If the growth is mainly limited by particle
diffusion through the background gas, however, then $\alpha _{meas}\left(
\sigma _{\infty }\right) \approx \alpha _{diff}.$

\begin{figure}[ht] 
  \centering
  \includegraphics[width=4in,keepaspectratio]{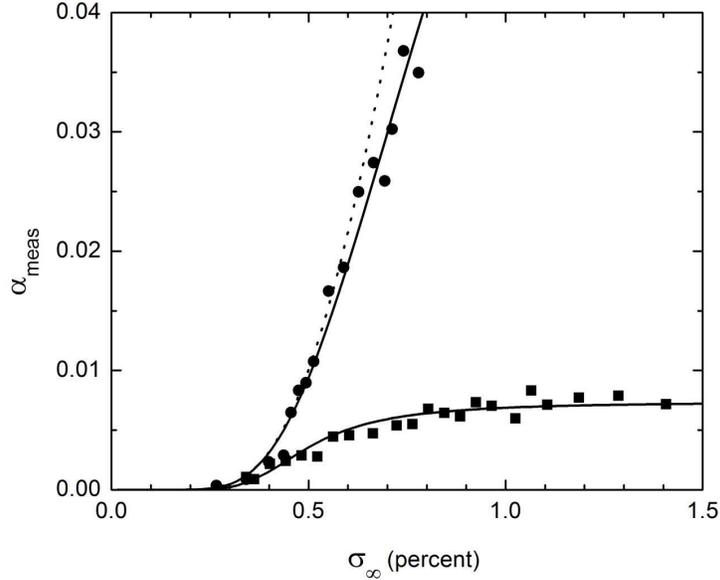}
  \caption{Measurements of the growth of
the basal facets of two ice crystals at -15 C, shown as the effective
condensation coefficient $\protect\alpha _{meas}$ as a function of
supersaturation $\protect\sigma _{\infty }$ far from the crystal. One
crystal (dots) was grown in a background pressure of air at 25 mbar, and the
other (squares) was grown in a background pressure of one bar. The
low-pressure crystal shows mainly kinetics-limited growth, while the growth
at high pressure is mainly limited by particle diffusion when the
supersaturation is high. Fit lines are described in the text. In addition,
the dotted line shows the derived intrinsic attachment coefficient $\protect%
\alpha _{intrinsic}(\protect\sigma _{surf})$.}
  \label{basalgrowth}
\end{figure}

Examples showing the growth of the basal facets of two ice crystals are
shown in Figure \ref{basalgrowth}, where the measured growth velocities have
been converted to $\alpha _{meas}\left( \sigma _{\infty }\right) $. At low
background pressures, we see that the growth is predominantly limited by
attachment kinetics, at least for small crystals at low supersaturations.
For essentially all our data, we found that the growth was well described by
a layer nucleation model \cite{saito}, and to describe the growth we adopted
a simplified parameterization of the intrinsic attachment coefficient $%
\alpha _{intrinsic}(\sigma _{surf},T)=A\exp (-\sigma _{0}/\sigma _{surf})$
where $A$ and $\sigma _{0}$ are parameters that depend on temperature but
not on supersaturation, and $\sigma _{surf}$ is the supersaturation at the
crystal surface.

At high background air pressures and high supersaturations, the growth
becomes distorted by particle diffusion effects. In Figure \ref{basalgrowth}%
, the solid lines show fits with $(A,\sigma _{0},\alpha _{diff})$ = $%
(1,2.3,0.15)$ and $(1,2.5,0.0075)$ for the low-pressure and high-pressure
crystals, respectively (with supersaturations in percent). The intrinsic
attachment coefficient $\alpha _{intrinsic}(\sigma _{surf})=\exp
(-2.3/\sigma _{surf})$ is also plotted as a dotted line in this figure. Note
that we have displayed $\alpha _{meas}(\sigma _{\infty })$ and $\alpha
_{intrinsic}(\sigma _{surf})$ in the same plot frame in Figure \ref%
{basalgrowth}, even though these can be significantly different quantities.
Our measurements directly yield $\alpha _{meas}(\sigma _{\infty })$, while
our goal is to extract $\alpha _{intrinsic}(\sigma _{surf})$ from these
measurements.

We explored the hypothesis that $\alpha _{intrinsic}$ may itself depend on
pressure, as the background gas could affect the surface molecular dynamics
that determines $\alpha _{intrinsic}.$ If the gas contains chemically active
components, then these will adsorb on the ice surface and affect its growth,
as is well known for the case of chemical impurities in air \cite{chemical}.

For pure gases such as nitrogen and oxygen, however, as well as clean air,
it is generally assumed that $\alpha _{intrinsic}$ does not depend on the
gas pressure. Comparisons of growth data taken in air at one bar and at 25
mbar, like the examples shown in Figure \ref{basalgrowth}, support this
conclusion. While the diffusion effects depend strongly on pressure, we
found no evidence that $\alpha _{intrinsic}$ was changed by the presence of
the background gas. We therefore made the implicit assumption in our data
that measurements of $\alpha _{intrinsic}$ made in clean air are equivalent
(after accounting for diffusion effects) to the ideal case of ice growth
from a gas of pure water vapor.

Although the diffusion distortions are relatively small at the low
background pressures used, we nevertheless found it important to correct for
these effects, and a detailed description of our correction procedure is
described in \cite{substrateinteractions}. Essentially, the correction
produced an estimate of $\sigma _{surf}$ using the known $\sigma _{\infty }$
along with the air pressure and measured growth rates of all the facets as
inputs. From this, the intrinsic attachment coefficient was then obtained
from $v=\alpha _{intrinsic}v_{kin}\sigma _{surf}.$

\begin{figure}[ht] 
  \centering
  \includegraphics[width=3.3in,keepaspectratio]{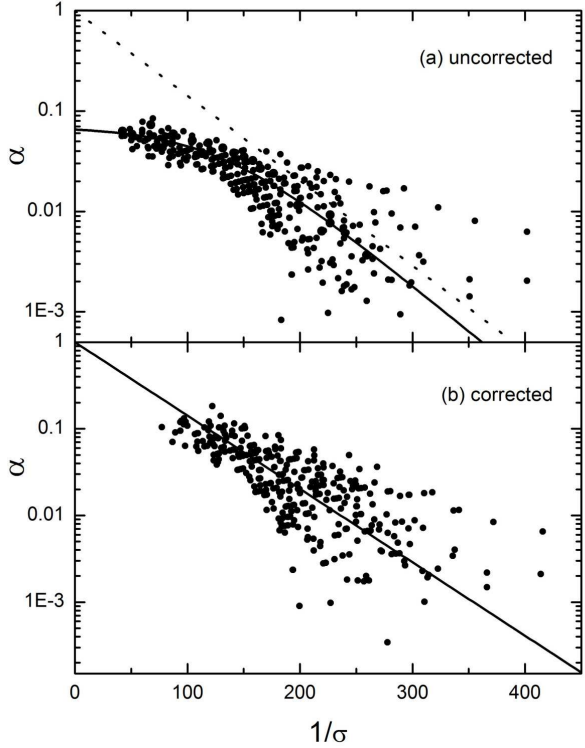}
  \caption{(a) A plot of the measured $%
\protect\alpha _{meas}(\protect\sigma _{\infty })$ of 21 ice crystals at $%
T=-12$ C. The solid line gives a fit $\protect\alpha _{meas}=\protect\alpha %
_{intrinsic}\protect\alpha _{diff}/(\protect\alpha _{intrinsic}+\protect%
\alpha _{diff}),$ where $\protect\alpha _{intrinsic}=\exp (-0.021/\protect%
\sigma )$ and $\protect\alpha _{diff}=0.07.$ (b) A plot of $\protect\alpha %
_{intrinsic}(\protect\sigma _{surf})$ for the same crystals after correcting
for diffusion effects. The solid line gives the fit $\protect\alpha %
_{intrinsic}=\exp (-0.0195/\protect\sigma ),$ and this same function is
shown as a dashed line in (a). Note that the overall horizontal shift in the
corrected points reflects the fact that $\protect\sigma _{surface}<\protect%
\sigma _{\infty }.$ }
  \label{oneover12}
\end{figure}

This data correction procedure allowed us to remove the residual diffusion
effects and convert $\alpha _{meas}(\sigma _{\infty })$ to $\alpha
_{intrinsic}(\sigma _{surf})$. Figure \ref{oneover12} shows a set of basal
growth data taken at $T=-12$ C, both without (top) and with (bottom) the
diffusion correction. Note that no adjustable parameters were use in this
conversion.

Plotting $\alpha $ as a function of $\sigma ^{-1}$ as in Figure \ref%
{oneover12}b shows how the parameters $A$ and $\sigma _{0}$ are coupled in
our measurements. The slope in Figure \ref{oneover12}b gives $\sigma _{0},$
while the intercept at $\sigma ^{-1}=0$ is equal to $A.$ Correcting for
diffusion effects was essential for determining $A$ accurately, since this
parameter was necessarily arrived at from an extrapolation of the data to $%
\sigma ^{-1}=0.$ For all our basal growth measurements, and for our prism
growth measurements with $T\leq -10$ C, the data were well described using $%
A=1,$ as shown in the example in Figure \ref{oneover12}. In these cases we
set $A=1$ and fit for $\sigma _{0}$ only, which reduced the uncertainty in
the $\sigma _{0}$ determinations.

\begin{figure}[t] 
  \centering
  \includegraphics[width=3.6in,keepaspectratio]{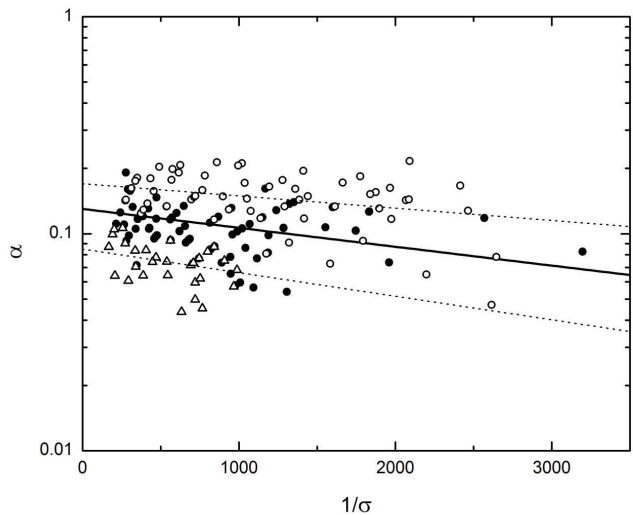}
  \caption{Measurements of $\protect\alpha %
_{intrinsic}(\protect\sigma _{surf})$ versus $\protect\sigma _{surf}^{-1}$
obtained from 19 crystals growing at a temperature of $-3$ C. The three
different sets of symbols are from separate measurement runs done on
different days. The lines show $\protect\alpha =A\exp (-\protect\sigma _{0}/%
\protect\sigma )$ with parameters $(A,\protect\sigma _{0})=$ (0.17,
0.00013), (0.13,0.0002), and (0.085, 0.00025).}
  \label{data3}
\end{figure}

For prism growth measurements at temperatures $T>-10$ C, it became clear
that $A=1$ would no longer fit the observations. Figure \ref{data3} shows
one example of corrected growth data at $T=-3$ C. For these measurements we
fit both $A$ and $\sigma _{0}$ to the data. We found that $\sigma _{0}$ was
low with these data, so coupling between the parameters was not a serious
problem, allowing both parameters to be measured reliably.

In addition to a much lower $A,$ Figure \ref{data3} also demonstrates small
systematic variations between three different data sets acquired during
three separate measurement runs. The high and low runs (open points) were
done while testing different substrate surface treatments, while the central
run (solid points) was done with a clean sapphire substrate. We suspect that
the variations seen in Figure \ref{data3} resulted mainly from residual
chemical vapors in the test chamber, associated with the surface treatments,
depositing on the top ice surface and affecting its growth. Nearly all our
data were taken using a substrate that was thoroughly cleaned before each
run, with no additional surface treatments.

Residual chemical impurities of this nature are an unavoidable uncertainty
in all ice growth experiments, since one cannot be certain how clean is
clean enough. Furthermore we have seen situations were chemical impurities
blocked the surface growth and thus reduced $\alpha _{intrinsic}$, as well
as circumstances where impurities reduced the nucleation barrier and thus
increased $\alpha _{intrinsic}$. We have used quantitative experiments to
examine how surface chemistry can affect ice crystal growth \cite{chemical},
and we believe that the current experiment is sufficiently clean that
residual chemical effects are not a serious systematic problem. The final
error bars in our measurements were adjusted to account for residual
systematic errors of this kind, as estimated empirically from run-to-run
variations seen throughout all our data.

\begin{figure}[ht] 
  \centering
  \includegraphics[width=3.4in,keepaspectratio]{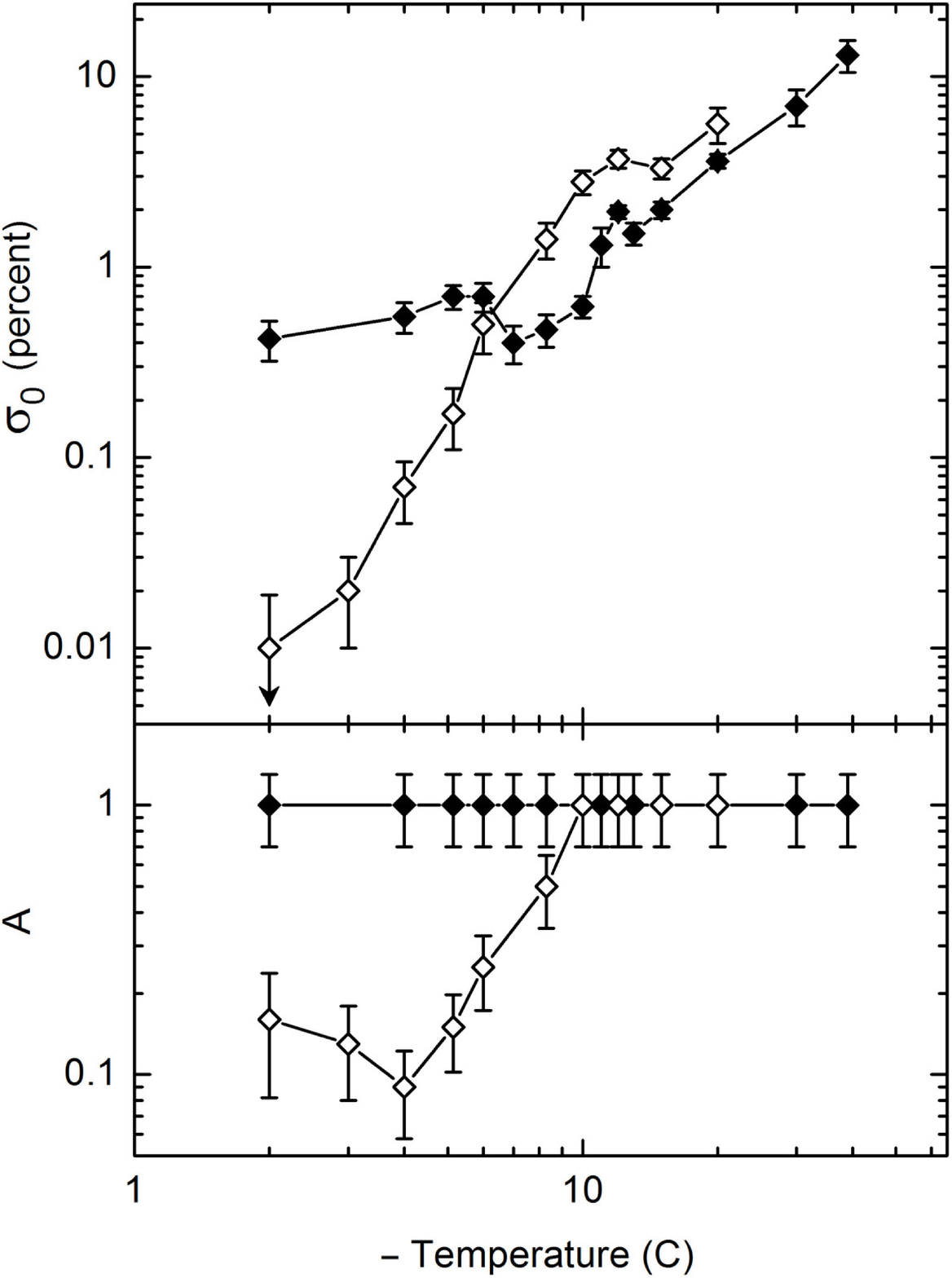}
  \caption{Measurements of the growth
behavior of the basal (solid points) and prism (open points) facet surfaces
of ice crystals. The intrinsic attachment coefficient was parameterized by $%
\protect\alpha _{intrinsic}(\protect\sigma _{surf},T)=A(T)\exp (-\protect%
\sigma _{0}(T)/\protect\sigma _{surf}),$ and the plots show the parameters $%
A(T)$ and $\protect\sigma _{0}(T)$ extracted from our data for both
principal facets.}
  \label{1combo}
\end{figure}

\begin{figure}[ht] 
  \centering
  \includegraphics[width=3.5in,keepaspectratio]{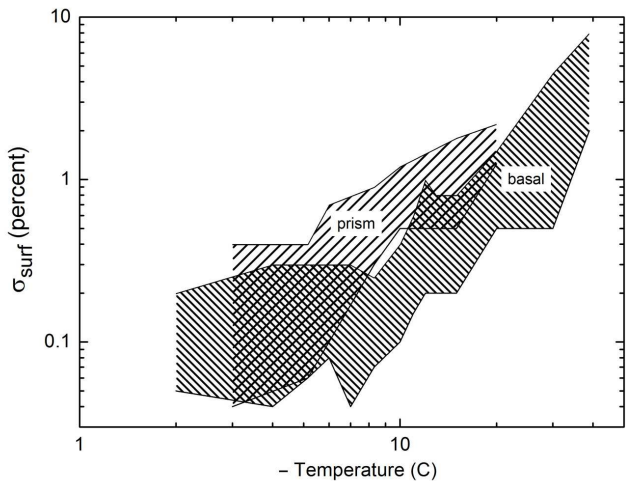}
  \caption{The ranges in $\protect\sigma _{surf}$
over which growth data were collected for the two facets. The ranges were
set mainly by the minimum and maximum growth velocities we could reliably
observe with this experiment.}
  \label{ranges}
\end{figure}

For both ice facets, and over the entire temperature range measured, we
found that our data are well described by the functional form $\alpha
_{intrinsic}(\sigma _{surf},T)=A\exp (-\sigma _{0}/\sigma _{surf}),$
indicative of a dislocation-free layer-nucleation model. From measurements
of over 200 crystals, we produced the final measurements of $\sigma _{0}(T)$
and $A(T)$ shown in Figure \ref{1combo}, which is the principal result from
this experiment. The current set of measurements includes only data points
for $T\geq -20$ C; lower-temperature points in Figure \ref{1combo} were
taken from \cite{oldgrowth}, which was a previous version of this
experiment. For those cases where we assumed $A=1$ in our fits, as described
above, we assigned error bars to $A$ in Figure \ref{1combo} that give an
estimate of the overall experimental uncertainty in this assumed value$.$The
ranges of surface supersaturations over which data were collected for these
measurements are shown in Figure \ref{ranges}.

Roughly 5-10 percent of the crystals sampled grew much more rapidly than the
norm, especially at low supersaturations, suggesting that the usual
nucleation barrier was substantially reduced \cite{substrateinteractions,
precisiongrowth}. We suspect that dislocations or perhaps isolated surface
chemical impurities affected the growth of these crystals, and they were
discarded from our data set before fitting to produce the measurements in
Figure \ref{1combo}.

Another set of crystals, again roughly 5-10 percent of those sampled, grow
anomalously slowly, and these crystals were also discarded from our fits. In
essence, we performed \textquotedblleft robust\textquotedblright\ fits to
our data -- first fitting the entire data set, then removing a small number
of \textquotedblleft outlier\textquotedblright\ crystals before redoing the
fits. This step was necessary because the outlier crystals grew
substantially differently from the norm and thus adversely affected the fits.

We have witnessed similar outlier effects in other ice growth experiments,
suggesting that this is a persistent and somewhat uncontrollable problem.
Managing this problem was possible only by measuring a large number of
crystals to define the overall distribution of growth behaviors, thus
allowing us to identify and remove highly unusual cases. We believe that the
final results in Figure \ref{1combo} accurately represent the growth of
chemically clean, dislocation-free ice facet surfaces.

\begin{figure}[ht] 
  \centering
  \includegraphics[width=3.9in,keepaspectratio]{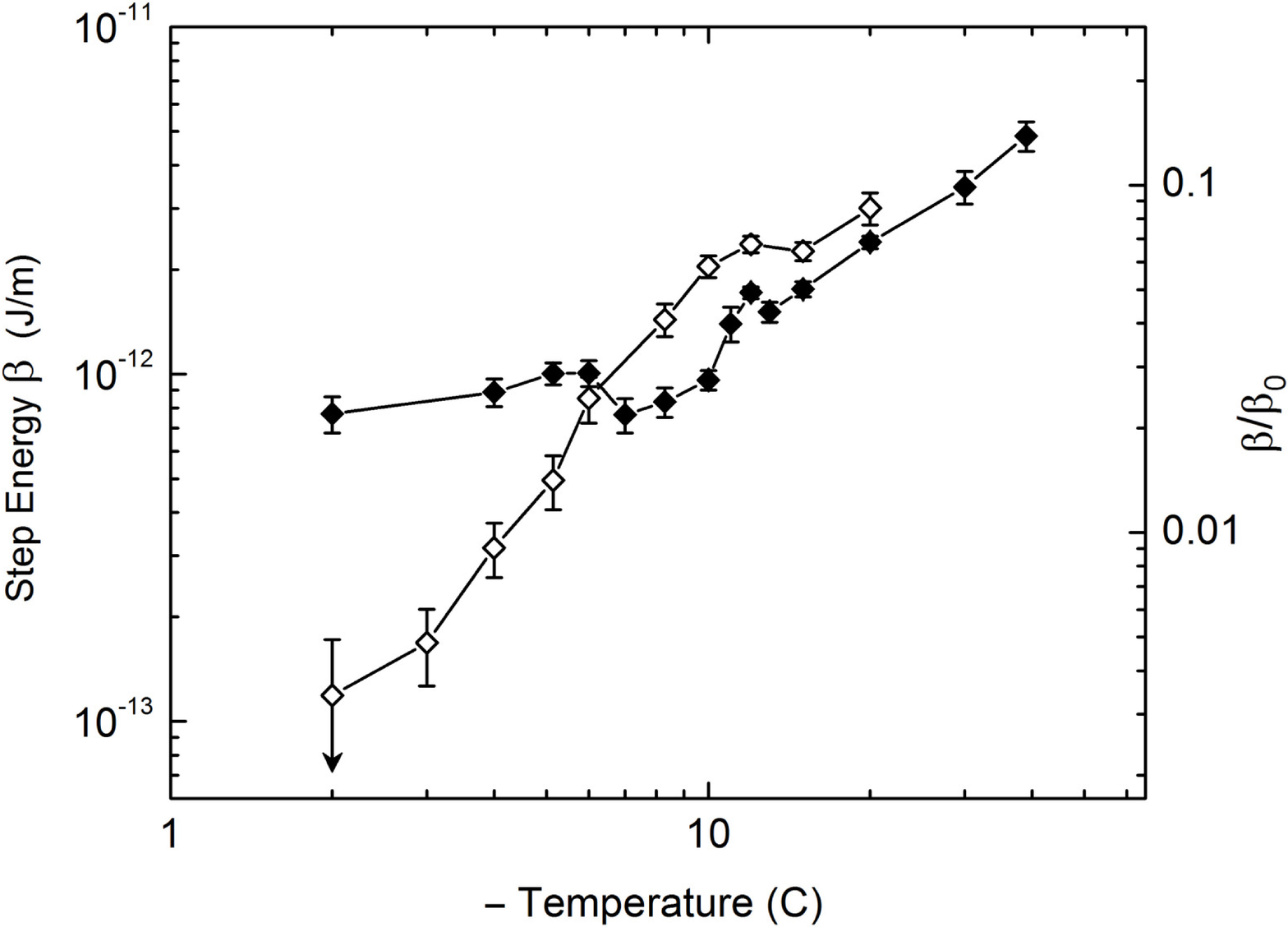}
  \caption{The step energy $%
\protect\beta (T)$ extracted from our measurements of $\protect\sigma %
_{0}(T) $ using classical nucleation theory, for the basal (solid points)
and prism (open points) facet surfaces. The scale on the right compares $%
\protect\beta $ with $\protect\beta _{0}=\protect\gamma a,$ where $\protect%
\gamma $ is the surface energy and $a$ is the step height.}
  \label{betagraph}
\end{figure}

Because the crystal growth we observed was everywhere well described by a
layer-nucleation model, the measured supersaturation parameter $\sigma _{0}$
can be used to calculate the terrace step energy $\beta (T)$ as a function
of temperature for both facet surfaces using 
\[
\sigma _{0}=\frac{\pi \beta ^{2}\Omega _{2}}{3k^{2}T^{2}} 
\]%
where $\Omega _{2}$ is the area of a molecule on the surface. This relation
comes from classical 2D nucleation theory \cite{saito, oldgrowth}. A plot of 
$\beta (T)$ from our data is shown in Figure \ref{betagraph}. We note from
the scale on the right side of Figure \ref{betagraph} that $\beta (T)$ is
much smaller than $\beta _{0}=\gamma a\approx 3.5\times 10^{-11}$ J/m, the
product of the surface energy $\gamma \approx 0.11$ J/m$^{2}$ of the
ice/vapor interface and the nominal molecular step height $a\simeq 0.32$ nm,
which is an upper limit on the step energy \cite{oldgrowth}.

\section{Discussion and Interpretation}

A comparison of the results from the present experiment with analogous
results from earlier ice growth experiments is beneficial for advancing the
state-of-the-art for these measurements. A careful scrutiny of the published
results reveals that in many older experiments a number of systematic errors
were not adequately managed (for a summary, see \cite{libbrechtreview,
critical}). Diffusion effects especially distort the ice growth behavior,
and this fact was not always adequately appreciated in prior experiments.

Removing diffusion effects to reliably determine $\alpha _{intrinsic}$ is
especially difficult in experiments done at pressures near one bar. Precise
diffusion modeling would be necessary to extract $\alpha _{intrinsic},$ and
we believe that sufficiently accurate modeling techniques have not yet been
demonstrated. Furthermore, our analysis shows that diffusion effects are
important even at quite low pressures, and that the corrections become
larger with larger crystals and at higher supersaturations. By identifying,
reducing, and modeling this and other systematic effects, we believe that
the current measurements are a substantial improvement over previous
attempts to determine $\alpha _{intrinsic}(\sigma _{surf},T).$

Interpreting our results, summarized by the parameterization of $\alpha
_{intrinsic}(\sigma _{surf},T)$ shown in Figure \ref{1combo}, presents a
significant challenge. Our theoretical understanding of the surface
structural changes that accompany surface melting is itself rather poor. How
these structural changes in turn affect the crystal growth dynamics is
clearly a complex many-body problem. We offer the following conclusions and
observations:

1) Our first conclusion, as stated above, is that the measured attachment
coefficients for both principal facets are well described by a
dislocation-free layer-nucleation model with the simplified parameterization 
$\alpha _{intrinsic}(\sigma _{surf},T)=A\exp (-\sigma _{0}/\sigma _{surf}).$
We find it quite remarkable that for both facets the growth dynamics of can
be summed up so concisely by the functions $A(T)$ and $\sigma _{0}(T).$ This
is true even going through the transition from essentially no significant
surface melting to a fully developed quasiliquid layer. Although the
equilibrium structure of the ice surface changes dramatically over this
temperature range, as does the equilibrium vapor pressure, the functional
form $\alpha \approx A\exp (-\sigma _{0}/\sigma )$ remains unchanged as the
ice growth is everywhere described by a layer nucleation model. The dominant
change with temperature is seen in the nucleation parameter $\sigma _{0}(T)$%
, accompanied by a relatively modest change in $A(T)$ on the prism facet.

2) The measured $\sigma _{0}(T)$ immediately yields the terrace step energy $%
\beta (T)$ from an application of classical 2D nucleation theory, as
described above. We note that the step energy is a fundamental property of
the ice surface, in much the same way that the surface energy is a
fundamental quantity. It is also an equilibrium property, even though it was
derived here from the dynamical process of crystal growth. As an
equilibrium, molecular-scale quantity, the step energy $\beta (T)$ may be
amenable to calculation using perturbation techniques or molecular dynamics
simulations. Considerable progress has been made in investigations of ice
surface melting using molecular dynamics simulations \cite{moldymice1,
moldymice2}, so perhaps step energies can be calculated using similar
methods. The observed strong temperature dependence in $\beta (T)$ may thus
yield important insights into how surface melting affects the ice surface
structure, in particular the interface between the crystalline solid and the
quasi-liquid surface layers. The step energy can also be used to infer some
features of the equilibrium crystal shape, which to date has not been
reliably measured for ice \cite{ecs}.

3) The growth behaviors of the basal and prism facets are remarkably
different, especially at temperatures above $T\approx -5$ C. In particular,
at the highest temperatures we see $\sigma _{0,basal}\gg \sigma _{0,prism},$
in a temperature regime where we would expect to find a thick quasi-liquid
layer on the ice surface. This is consistent with the growth behavior of ice
from liquid water, where the basal surface grows much more slowly than the
prism surface at low supercoolings. Quantitatively relating the crystal
growth rates from water vapor and from liquid water may be a tractable
theoretical problem, since the surface attachment kinetics is likely similar
for these two cases. To our knowledge, however, the kinetic coefficient for
ice growth from liquid water has not yet been determined, as growth
measurements are usually limited by heat diffusion effects.

4) For both facets we see an overall trend that $\beta (T)$ decreases with
increasing $T$. As the temperature increases, we suggest that surface
restructuring smooths out the terrace edge, thereby lowering the step
energy. The trend with temperature is particularly simple on the basal facet
for $T\leq -15$ C, where we expect the complicating effects from surface
melting are small. These low temperatures may be best suited for molecular
dynamics simulations, so extending the prism facet data to similarly low
temperature would be beneficial.

5) The small bump in $\sigma _{0,basal}(T)$ at $T\approx -12$ C is a
significant and robust feature in our data, and we suggest that this feature
identifies the onset of significant surface melting (significant with regard
to its effect on crystal growth dynamics) on the (0001) surface at this
temperature. The mechanism that might produce this bump, however, is
unknown. This onset temperature agrees with that measured by \cite{dosch},
which is a surface melting measurement using an ice surface preparation
similar to that used in the current experiment. The observed shoulder in $%
\sigma _{0,prism}(T)$ near $-12$ C may be a related phenomenon. More
generally, we see that the behavior of $\sigma _{0}(T)$ for both facets is
most complex in the temperature range $-15$ C $<T<-5$ C, and we suggest that
this range includes the onset and development of surface melting on both
facets. At lower temperatures surface melting is largely absent, while at
higher temperatures the quasi-liquid layer is fully developed.

6) The change in $\sigma _{0,basal}(T)$ near $T=-5.5$ C is especially
noteworthy, as it reverses the general trend with temperature seen in $%
\sigma _{0}(T)$ for both facets. One possibility is that surface melting
initially results in a more disordered basal surface with a reduced step
energy, producing the dip in $\sigma _{0,basal}(T)$ near $T=-7$ C. At higher
temperatures, however, we suggest that the quasi-liquid layer (QLL) becomes
more fully developed, producing a relatively sharp QLL/ice interface. This
sharper interface may then result in a more distinct terrace step and a
higher step energy.

7) We note that our measurements of $\alpha _{intrinsic}(\sigma _{surf},T)$
do not lead one immediately to an explanation of the well-known morphology
diagram describing ice growth from water vapor \cite{libbrechtreview}. In
fact, the opposite is true. At $-15$ C, for example, we find $\alpha
_{basal}>a_{prism}$ at all $\sigma ,$ which is at odds with the occurrence
of very thin plate-like crystals at this temperature. The explanation for
this and other morphological discrepancies seems to lie in the fact that our
measurements give $\alpha _{intrinsic}(\sigma _{surf},T)$ only for flat
facet surfaces. The edge of a thin plate, however, is not a flat facet
surface. The molecularly flat prism facet in this case is typically only
some hundreds of molecules wide (assuming an edge radius of curvature of
approximately one micron), and it is the dynamics on this surface that
determines the edge growth velocity. Earlier one of us put forth the
hypothesis of \textit{structure-dependent attachment kinetics}, suggesting
that $\alpha $ changes depending on the local crystal morphology, and in
particular that $\sigma _{0}$ is reduced on thin edges \cite{sdak}. We
believe that including this hypothesis can connect the present measurements
with the morphology diagram, although many details remain unknown at present.

In summary, we have measured the ice crystal growth velocities of the
principal facets of ice from water vapor for what approaches the ideal case
-- that of perfect crystalline surfaces growing in near equilibrium with
pure water vapor. We find that data such as these can be used both as a
probe of the temperature dependence of surface melting and as a measure of
the effects of surface melting on crystal growth dynamics. This work was
supported in part by the California Institute of Technology and the
Caltech-Cambridge Exchange (CamSURF) program.

\end{document}